\documentclass[aps,prb,showpacs,twocolumn,superscriptaddress]{revtex4}
\usepackage{amsmath}
\usepackage{amsfonts}
\usepackage{amssymb}
\usepackage{graphicx}

\begin{document}

\title{ Multi-orbital Cluster Perturbation Theory for transition metal oxides }

\author{F.~Manghi}

\address{Dipartimento di Fisica, Universit\`a di Modena e Reggio Emilia and CNR - Institute of NanoSciences - S3, Via Campi 213/A, I-41125 Modena, Italy }

\begin{abstract}

We present an   extension of Cluster Perturbation Theory  to  include many body correlations associated to local e-e repulsion in real materials. We show that this approach can  describe the physics of complex correlated materials where different atomic species and different orbitals coexist. The prototypical case of MnO is considered.

\end{abstract}

\pacs{71.30.+h, 71.27.+a, 71.20.Be}
\maketitle

The competition between inter-site hopping and on-site electron-electron repulsion dominates the physics of transition metal oxides \cite{RevModPhys.70.1039}. Standard band theory based on the independent particle approach predicts these large gap insulators to be metallic in the paramagnetic phase and fails in reproducing the band width and satellite structures observed in the experiments. Only approaches that augment band theory with true many body effects such as 3-Body Scattering theory (3BS) \cite{ManghiNiO,ManghiNi,ManghiCo} and Dynamical Mean Field Theory (DMFT) \cite{Dimarco_NiO,Anisimov_NiO} have been able to reproduce the band gap in the paramagnetic state and to describe photoemission data. However the agreement between experiments and  many-body calculations is still far from being fully quantitative  \cite{Barriga1,Barriga2,Barriga3} and different theoretical methods are constantly explored.

In this paper we show that a multi-orbital extension of Cluster Perturbation Theory (CPT) \cite{Senechal3} can be applied to the study of quasi-particle excitations in transition metal monoxides. CPT solves the problem of many interacting electrons in an extended lattice by approaching first the many body problem in a subsystem of finite size - a cluster-  and then embedding it within  the infinite medium. CPT shares this strategy with other approaches such as Variational Cluster Approach (VCA)\cite{VCA,SFA} and Cellular Dynamical Mean Field Theory \cite{Senechal_CDMFT} where the embedding procedure is  variationally optimized.

We use here MnO as a test case. We restrict to  the paramagnetic phase  at zero pressure where, according to single particle  band structure, MnO is  metallic with  half occupied  d-orbitals - a paradigmatic case to study Mott-Hubbard metal-to-insulator transition. \footnote{For an account of the  rich phase diagrams of MnO as a function of temperature and pressure see references  \cite{PhysRevLett.94.115502,PhysRevB.81.115116,Kunes}.}

The paper is organized as follows: in section  \ref{one} we recall the CPT theory and outline its extension to the many-orbital case; in section \ref{two} we describe how the cluster Green function is calculated in a complex lattice with more than one atomic species and  many orbital per site;  section \ref{three} is for the discussion of the results obtained for MnO.

\section{Multi-orbital CPT}
\label{one}

In CPT  the lattice is  seen as the periodic repetition of identical clusters (Fig. \ref{steps} ) and the Hubbard Hamiltonian  can  be
partitioned in two terms, an intra-cluster ($ \hat{H}_c$) and an inter-cluster one ($\hat{V} $)
\begin{equation}
\label{hubbard1}
\hat{H}  = \hat{H}_c+\hat{V}
\end{equation}
where
\begin{eqnarray}
\label{hubbard2}
\hat{H}_c &=& \sum_{i l \alpha} \epsilon_{i l\alpha} \hat{n}_{i l\alpha}+ \sum_{\alpha \beta } \sum_{ij l }
t_{i l \alpha, j l\beta} \hat{c}_{i l\alpha }^{\dagger}
\hat{c}_{j l \beta } \nonumber \\
&+&\sum_{i l \alpha \beta}  U^i_{\alpha \beta}
 \hat{n}_{i l\alpha\uparrow} \hat{n}_{i l\beta \downarrow} \nonumber \\
\hat{V} &=& \sum_{\alpha \beta } \sum_{ij l\neq l'}
t_{i l \alpha, j l'\beta} \hat{c}_{i l\alpha }^{\dagger}
\hat{c}_{j l' \beta }
\end{eqnarray}

Here $\alpha, \beta$ are orbital indexes, $ \epsilon_{i l\alpha}$ are intra-atomic orbital parameters and $t_{i l \alpha, j l'\beta}$ hopping terms connecting orbitals centered on different sites.  Each atom  is identified  by the cluster it belongs to (index $l$) and by its position inside the cluster (index $i$). The lattice is a collection of $L \to \infty$ clusters each of them containing M atoms whose position is identified by the vector  $\textbf{R}_l$+$\textbf{r}_i$. Each  atom in the cluster is characterized by a set of  orbitals $n_i^{orb}$ and $ K=\sum_{i=1}^M n_i^{orb}$ is the total number of sites/orbitals per cluster.

Since in the Hubbard model the e-e Coulomb interaction is on-site, the inter-cluster hamiltonian $\hat{V} $ contains only single particle terms,  the  many body part being   present in the intra-cluster hamiltonian $ \hat{H}_{c}$ only, a key feature for the practical implementation of the method.
Having partitioned the Hamiltonian in this way an exact expression involving  the resolvent operator $ \hat{G} $ is obtained
\[\hat{G}^{-1}=  z-\hat{H}_c-\hat{V} = \hat{G^c}^{-1}-\hat{V}
\]
and from this
 \begin{equation}\label{Dyson}
\hat{G} =   \hat{G^c} + \hat{G^c} \hat{V} \hat{G}
 \end{equation}
The one-particle propagator
\begin{eqnarray}
 \mathcal{G}(\textbf{k} n  \omega)&=& <\Psi_0|\hat{c}_{\textbf{k}  n }^{\dag}\hat{G} \hat{c}_{\textbf{k} n }|\Psi_0> \\ \nonumber
& +& <\Psi_0|\hat{c}_{\textbf{k}  n }\hat{G} \hat{c}_{\textbf{k} n }^{\dag}|\Psi_0>
\end{eqnarray}
is obtained exploiting the transformation from Bloch to localized basis
\[
\hat{c}_{\textbf{k}  n}^{\dag}=\frac{1}{\sqrt{M}}\sum_{i l \alpha}\mathcal{C}^n_{i \alpha}(\textbf{k})^* e^{-i\textbf{k}\cdot(\textbf{R}_l+\textbf{r}_i)}\hat{c}_{i l \alpha}^{\dag}
\]
and similarly for $\hat{c}_{\textbf{k}  n  } $. Here   $n$ is a band index and $\mathcal{C}^n_{i \alpha}(\textbf{k})$ are the eigenstate coefficients obtained by a band calculation for a superlattice of $L$ identical clusters  and the summation is over  $M=L\times K$. We get
\begin{equation}\label{gk}
 \mathcal{G}(\textbf{k} n  \omega)= \frac{1}{K}\sum_{i i' \alpha \beta}
 e^{-i\textbf{k} \cdot (\textbf{r}_i - \textbf{r}_{i'})}\mathcal{C}^n_{i \alpha}(\textbf{k})^* \mathcal{C}^n_{i \beta}(\textbf{k}) \mathcal{G}_{i \alpha i' \beta}(\textbf{k}   \omega)
\end{equation}
where $\mathcal{G}_{i \alpha i' \beta}(\textbf{k}   \omega)$ is the superlattice Green function, namely the Fourier transform of the Green function in the local basis
\begin{equation}
\label{gk1}
\mathcal{G}_{i \alpha i' \beta}(\textbf{k}   \omega)=\frac{1}{L}\sum_{l l'}e^{-i\textbf{k}\cdot(\textbf{R}_l-\textbf{R}_{l'})}
\mathcal{G}^{l l'}_{i \alpha i' \beta}(  \omega)
\end{equation}
This is the quantity that can be calculated  by eq.\ref{Dyson} that  explicitely becomes:
\begin{equation}\label{QC}
\mathcal{G}_{i \alpha i' \beta}(\textbf{k}    \omega)={\mathcal{G}^c}_{ i \alpha i' \beta}(  \omega)+ \sum_{j \gamma} B_{i \alpha j \gamma}(\textbf{k}  \omega) \mathcal{G}_{j \gamma  i' \beta}(\textbf{k}   \omega)
\end{equation}
where the $K\times K$ matrix $B_{i \alpha j \gamma}(\textbf{k}   \omega)$
is the Fourier transform of $\hat{G^c}\hat{V}$ involving neighboring sites that belong to different clusters.

Once the cluster Green function in the local basis  ${\mathcal{G}^c}_{ i \alpha i' \beta}(  \omega)$ has been obtained    by exact diagonalization, eq. \ref{QC} is solved   by a $K\times K$  matrix inversion at each $\textbf{k}$ and $\omega$. The quasi particle spectrum is then obtained in terms of  spectral function $A(\textbf{k}  \omega)$
\begin{equation}
\label{akn}
A(\textbf{k}   \omega) = \frac{1}{\pi}\sum_n Im  \mathcal{G}(\textbf{k} n \omega).
\end{equation}

\section{Cluster calculation for TM oxides}
\label{two}

\begin{figure}
\includegraphics[width=8cm]{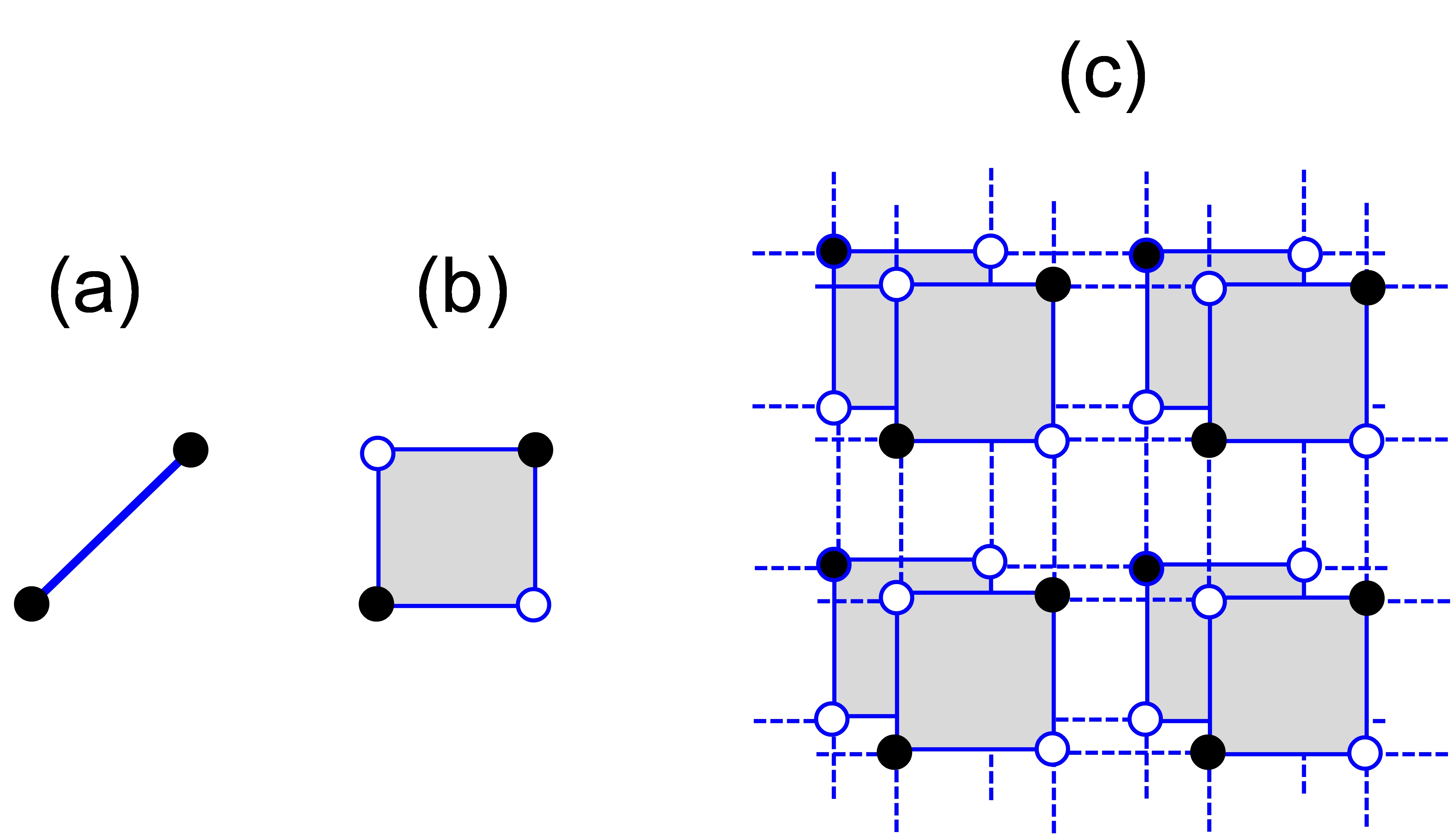}\\
  \caption{(Color on-line) Building blocks of the 3D Rocksalt structure for a transition metal mono-oxide: (a) a dimer of 2 TM atoms (filled black circles); (b) a $2 \times 2$  plaquette containing the two  atomic species (Oxygens as open circles); (c) stacking of  plaquette layers reproducing the 3D lattice. Dotted lines indicate the inter-cluster hopping.}\label{steps}
\end{figure}

The valence and first conduction   states of TM oxides  are  described by  TM \emph{spd}   and  oxygen \emph{sp} orbitals.  The dimer with $M=2$ TM atoms and $K=10$ \emph{d} orbitals (Fig. \ref{steps} a) is the basic unit where we will perform the exact diagonalization.

We recall that the exact diagonalization corresponds to write the manybody wavefunction as a superposition of    Slater determinants  that can be built by putting N electrons of spin up and N electrons of spin down on $K$ boxes:
\begin{equation}\label{EDPsi}
|\Phi_n^N> = \sum_l^{nconf} C_l^n |S_l>
\end{equation}
with
\begin{equation}
 |S_l>= \hat{c}^{\dag}_{l_1 \uparrow}\hat{c}^{\dag}_{l_2 \uparrow} ... \hat{c}^{\dag}_{l_{N} \uparrow}\hat{c}^{\dag}_{l_{N+1} \downarrow}\hat{c}^{\dag}_{l_{N+2} \downarrow} ... \hat{c}^{\dag}_{l_{N+N} \downarrow}|0>
\end{equation}
Each Mn atom brings to the dimer 5 $d$ electrons (half occupation) and the dimension of the Hilbert space spanned by the Slater determinants is $nconf=(\frac{K!}{N!(K-N)!)})^{2} = 63504$.
We separately solve the problem  with N, N-1 and N+1 electrons and
calculate the dimer Green function using the Lehmann representation, namely
\begin{eqnarray}
\label{dimer}
   \mathcal{G}^{dd}_{i \alpha i' \beta}&(&\omega) = \sum_n \frac{<\Phi_0^N|\hat{c}^{\dag}_{i \alpha } |\Phi_n^{N-1}><\Phi_n^{N-1}|\hat{c}_{i' \beta}|\Phi_0^N>}{\omega - (E_0^N- E_n^{N-1})}
 \nonumber \\
  &+&
   \sum_n \frac{<\Phi_0^N|\hat{c}_{i \alpha } |\Phi_n^{N+1}><\Phi_n^{N+1}|\hat{c}^{\dag}_{i' \beta } |\Phi_0^{N}>}{\omega - (E_n^{N+1}-E_0^N)}
\end{eqnarray}

Due to the large dimensions of the matrix to be diagonalized  the band-Lanczos algorithm \cite{Lanczos} is used to obtain $\sim 1000 $ eigenvalues and eigenvectors $E_n^{N\pm1}$, $\Phi_n^{N\pm1}$  for the system with $N\pm1$ electrons as well as the ground state $E_n^{N}$, $\Phi_0^{N}$ for N electron system.

The dimer problem that we have described accounts for both hopping and e-e repulsion on the $d$ orbitals of TM atoms and therefore includes a large part of the relevant physics of the interacting system. In particular, since the system is half occupied, we expect the ground state $E_0^{N+1}$ to be  larger than
$ E_0^{N-1}$ with an energy distance growing with $U$. This is promising in view of a gap opening in the extended system.

Notice however that this dimer  does not represent a  partition (in mathematical sense) of the 3D rocksalt lattice and therefore it is not the cluster to be used in the  CPT procedure described in the previous section. The smallest unit that has the necessary characteristics to reproduce without overlaps the 3D rocksalt lattice is the  $2 X 2$ plaquette of Fig. (\ref{steps} b ). It  contains  both TM atoms  and  oxygens and the  Hamiltonian $\hat{H}_c$ of equation \ref{hubbard1} is  a sum of on-site and inter-site terms connecting  TM \emph{d} orbitals (type A)  and \emph{sp} orbitals of both TM and oxygen atoms (type B):

\begin{eqnarray}
\label{hcluster}
\hat{H}_c &=&  \hat{H}_c^{diag}  + \hat{V}_c^{AB}
\end{eqnarray}
with
\begin{eqnarray}
\label{hubbard3}
\hat{V}_c^{AB} &=& \sum_{\alpha_A \beta_B }
t_{i l \alpha_A, j l\beta_B} \hat{c}_{i l\alpha_A }^{\dagger}
\hat{c}_{j l' \beta_B }
\\ \nonumber
\hat{H}_c^{diag} &=& \hat{H}_c^{AA} + \hat{H}_c^{BB}
\end{eqnarray}
where
\begin{eqnarray}
\hat{H}_c^{AA} &=& \sum_{i l \alpha_A} \epsilon_{i l\alpha_A} \hat{n}_{i l\alpha_A}+ \sum_{\alpha_A \beta_A } \sum_{ij l }
t_{i l \alpha_A, j l\beta_A} \hat{c}_{i l\alpha_A }^{\dagger}
\hat{c}_{j l \beta_A } \nonumber \\
&+&\sum_{i l \alpha_A \beta_A}  U^i_{\alpha_A \beta_A}
 \hat{n}_{i l\alpha_A\uparrow} \hat{n}_{i l\beta_A \downarrow}
\end{eqnarray}
and a similar expression for $\hat{H}_c^{BB}$.

We  need  therefore to \emph{embed} the dimer into the plaquette, in other words we need  to  write the cluster Green function in terms of the dimer one.  This can be done noticing again that
\[\hat{G^c}^{-1}=  z-\hat{H}_c = (\hat{G}^{diag})^{-1}-\hat{V}_c^{AB}. \]

that results as before in a Dyson-like equation
\begin{equation}
\label{Dyson1}
\hat{G^c}=  \hat{G}^{diag}  + \hat{G}^{diag} \hat{V}_c^{AB} \hat{G^c}
\end{equation}
In the local basis $\hat{G}^{diag}$ is block-diagonal and the non-zero elements $\hat{G}^{diag}_{AA}, \hat{G}^{diag}_{BB}$  are obtained by performing separate exact diagonalizations that include either A or B orbitals: $\hat{G}^{diag}_{AA}\equiv \hat{G}^{dd}$ is the dimer Green function of eq. \ref{dimer} while $\hat{G}^{diag}_{BB}$ involves only $sp$ orbitals and in the present case is non-interacting.  In the local basis eq. \ref{Dyson1}  can be solved by performing a matrix inversion.

\begin{equation}
\hat{G^c}=  \hat{G}^{diag}  \times (\hat{G}^{diag} \hat{V}_c^{AB} )^{-1}
\end{equation}
or more explicitely

\begin{equation}
\label{inv}
{\mathcal{G}^c}_{ i \alpha i' \beta}(  \omega)= \sum_{j \gamma }{\mathcal{G}^{diag}}_{ i \alpha j \gamma}(  \omega) \times ( \hat{G}^{diag} \hat{V}_c^{AB}  )^{-1}_{j \gamma i' \beta'}
\end{equation}
with indices running over  $K=26$ sites/orbitals of the plaquette  (9 $spd$ orbitals on 2 TM atoms and 4 $sp$ orbitals on 2 Oxygens).

We want to stress that the present formulation  is nothing else than the  extension of CPT to the case of more orbitals per site when it is necessary to deal with exceedingly large dimensions of the configuration space. The CPT prescriptions in this case may be rephrased as follows:  chose a  partition of the lattice Hamiltonian into a collection of non overlapping clusters connected by inter-cluster hopping; make a further partition inside each cluster defining a   suitable collections of   sites/orbitals; perform separate exact diagonalizations plus matrix inversion to calculate the cluster Green function in local basis  by eq. \ref{inv}  and finally obtain the full lattice Green function in a Bloch basis by adding the cluster-cluster hopping terms according to eq. \ref{QC}.

A final comment on the approximations involved: in the same way as in the standard single-orbital CPT, writing the lattice Green function  in terms of Green functions of decoupled subunits  amounts to  identify the many electron  states of the extended lattice as the product of cluster few electron ones. In the present case in particular, choosing the  TM dimer as the basic unit we have excluded from the few-electron eigenstates obtained by exact diagonalization the contribution of oxygen $p$ orbitals, treating the O $p$ - TM $d$ hybridization by the embedding procedure (eq. \ref{inv}) and by the periodization (eq. \ref{QC}). This approximation can be improved by some kind of variational procedure but in any case it interesting to assess its validity \emph{per se}, for instance  by comparing theory and experiments in specific cases. This is what we do in the next section.

\section{Application to $MnO$}
\label{three}

\begin{figure}
\includegraphics[width=9cm]{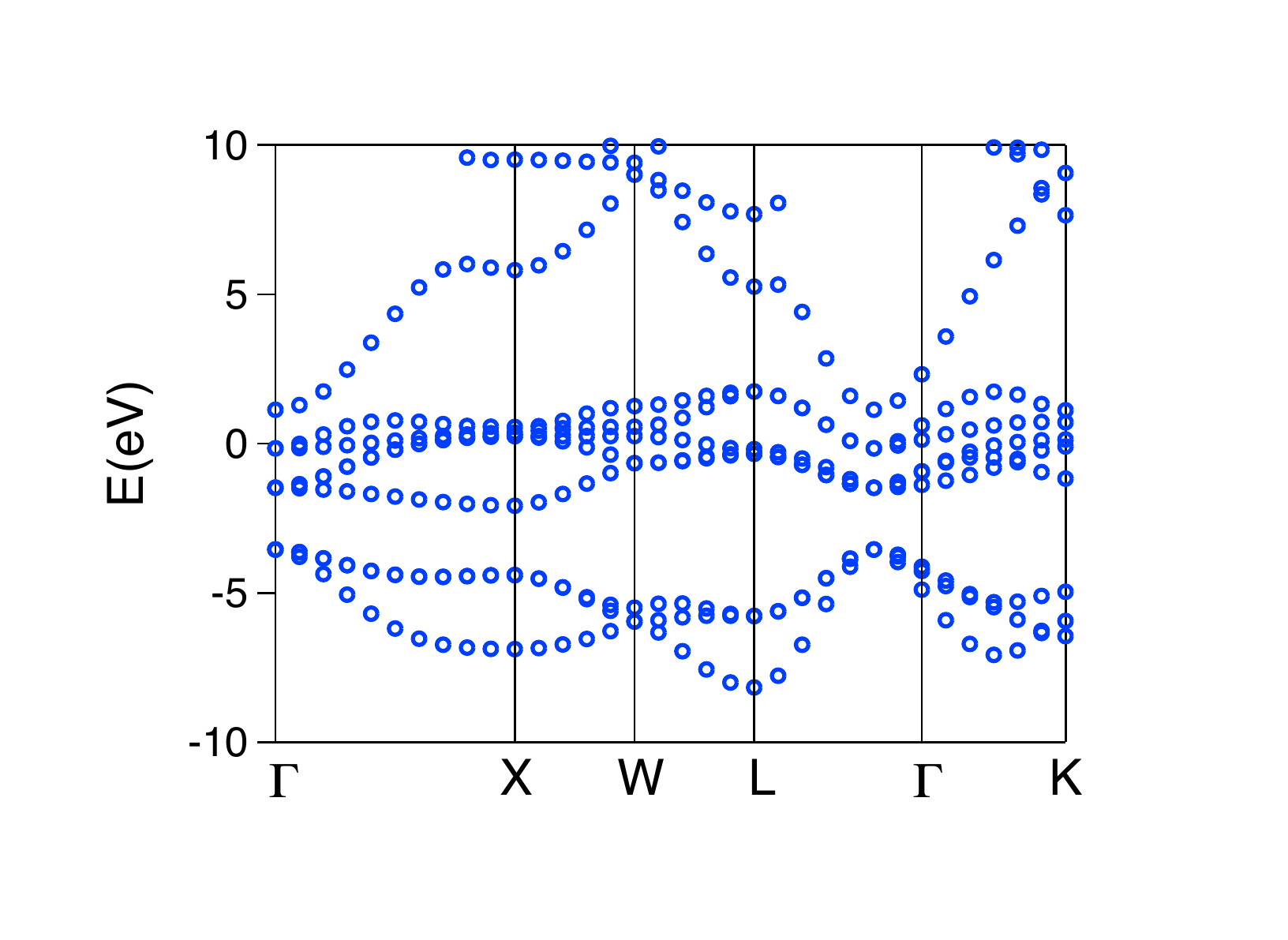}\\
  \caption{(Color on-line) Single particle band structure of MnO obtained with the Tight-Binding parametrization of Tables I, II .}\label{bande0}
\end{figure}

The non interacting contribution to the   Hubbard Hamiltonian of eq. \ref{hubbard1} can be written  as a standard Tight-Binding Hamiltonian  in terms of Koster-Slater \cite{KS} parameters  obtained by  a least squares fitting of an ab-initio band structure.  The parameters obtained by fitting the band structure of MnO calculated in the DFT-LMTO scheme \cite{LMTO} are reported in Tables I,II and give rise to the  band structure  of Fig. \ref{bande0}.

 \begin{table*}[t]
 \label{table0}
 \begin{center}
  \caption{\label{KS1} On site Koster-Slater parameters (in eV)
 for MnO. }
  \begin{tabular}{ccccccc}
    &$E_s(Mn)$ & $E_p(Mn)$ & $E_{t2g}(Mn)$ & $E_{eg}(Mn)$ & $E_s(O)$ & $E_p(O)$ \\
      \hline
 & 7.313 &11.546& -0.763  & -0.010 & -18.553  & -4.806\\
   \end{tabular}
 \end{center}
\end{table*}

 \begin{table*}[t]
 \label{table1}
 \begin{center}
  \caption{\label{KS2} Inter-site Koster-Slater parameters (in eV)
 for MnO . }
  \begin{tabular}{llcccccccccc}
  &   & $ss_{\sigma}$ & $pp_{\sigma}$ & $pp_{\pi}$ & $dd_{\sigma}$ & $dd_{\pi}$ & $dd_{\delta}$ & $sp_{\sigma}$ & $sd_{\sigma}$ & $pd_{\sigma}$ & $pd_{\pi}$ \\
  \hline
$Mn$ & $Mn$ & -0.514 & 1.435  & -0.137  & -0.353& 0.028 & 0.047 & 0.486 & -0.285 & -0.081  & 0.209 \\
$O$ & $Mn$  & 0.0    & 0.0    & 0.0     & 0.0   & 0.0   & 0.0   & 0.0   & -1.074 & -1.243  & 0.632 \\
$O$ & $O$   & -0.124 &  0.519 & -0.102  &  0.0  &  0.0  & 0.0   &-0.016 & 0.0    & 0.0     & 0.0 \\
   \end{tabular}
 \end{center}
\end{table*}

When using  TB parameters in the Hubbard Hamiltonian  we must take care of the double-counting issue: ab-initio band structure, and the TB parameters deduced from it,  contain  the e-e Coulomb repulsion as a mean-field  that must be  removed before including $U$ as a true many body term.  "Bare" on-site parameters should be calculated by subtracting the mean filed value of the Hubbard term, namely
\begin{equation}\label{double_counting}
E_{\alpha \sigma}^*=E_{\alpha} -\sum_{i } U^i_{\alpha} <n_{i \alpha -\sigma}>
\end{equation}
 This definition involves the $d$ occupation  inside the cluster  used in the exact diagonalization and cancels out the energy shift due to double-counting within each cluster. Notice that $<n_{i \alpha \sigma}>=<n_{i \alpha -\sigma}>$  and  $E_{\alpha \sigma}^*$ is spin-independent.

We tested our approach using different $U$ values  and we report the results obtained for $U=9 eV$ . This value optimizes the agreement between theory and experiments and is not far from the values reported in the literature ranging from U=6.0 up to U=8.8  \cite{Dimarco_NiO,PhysRevB.44.1530,PhysRevB.81.115116,Ferdi2013}.
Since we have ignored  the  orbital dependence of $U$ as well as  the e-e repulsion among parallel spins the present value  $U=9$ should be considered as an effective one.

\begin{figure}
\includegraphics[width=9cm]{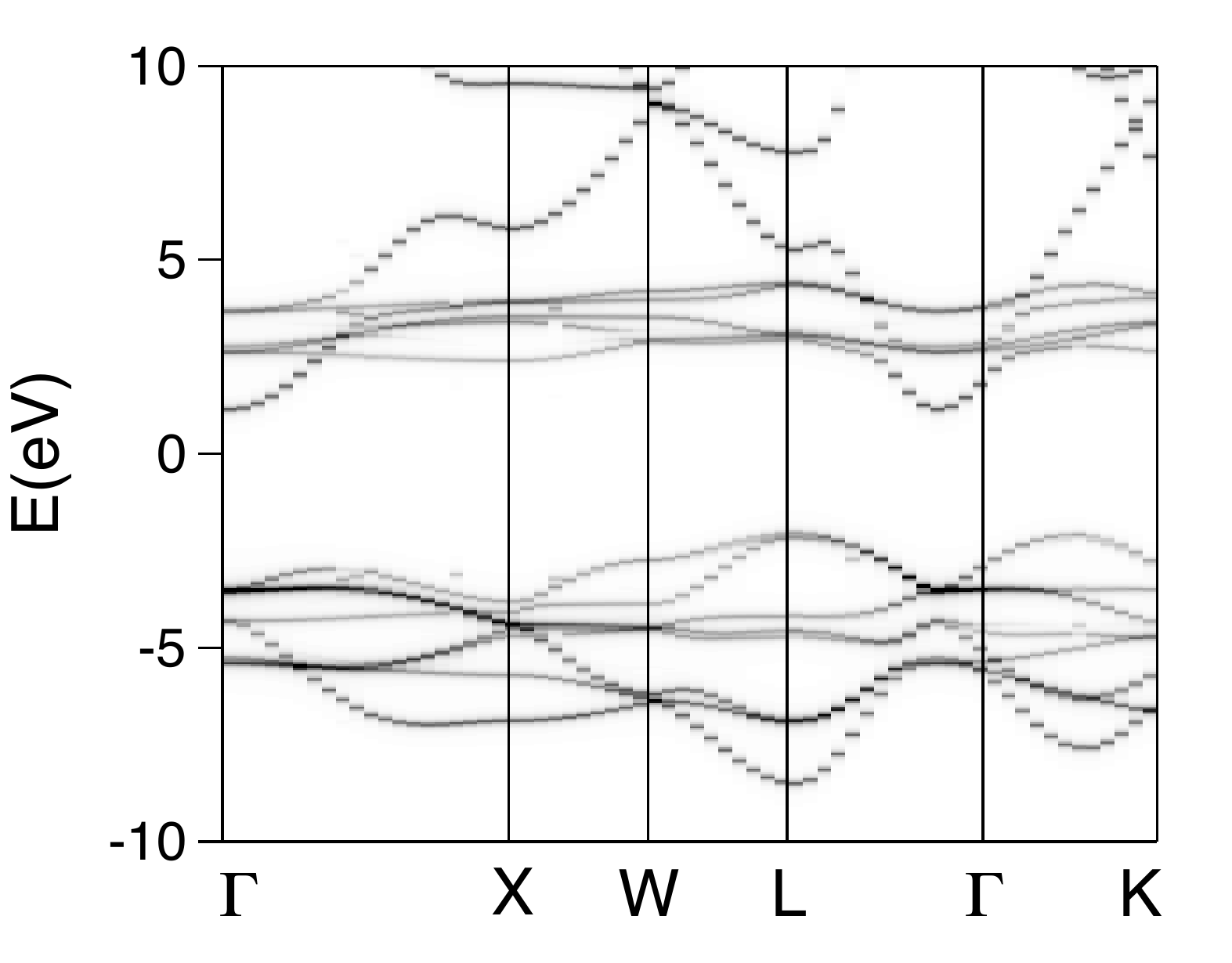}\\
  \caption{k-resolved spectral functions describing the quasi-particle band structure of MnO for  $U=9$.}\label{bandeU}
\end{figure}

The quasi-particle band structure of MnO is shown in Figure \ref{bandeU} where we plot the calculated k-resolved spectral function (eq. \ref{akn} ). We notice that the Mn $d$ band that in the  absence of correlation  (Fig. \ref{bande0}) crosses the Fermi level is now split in   lower and   upper Hubbard bands.

\begin{figure}
\includegraphics[width=9cm]{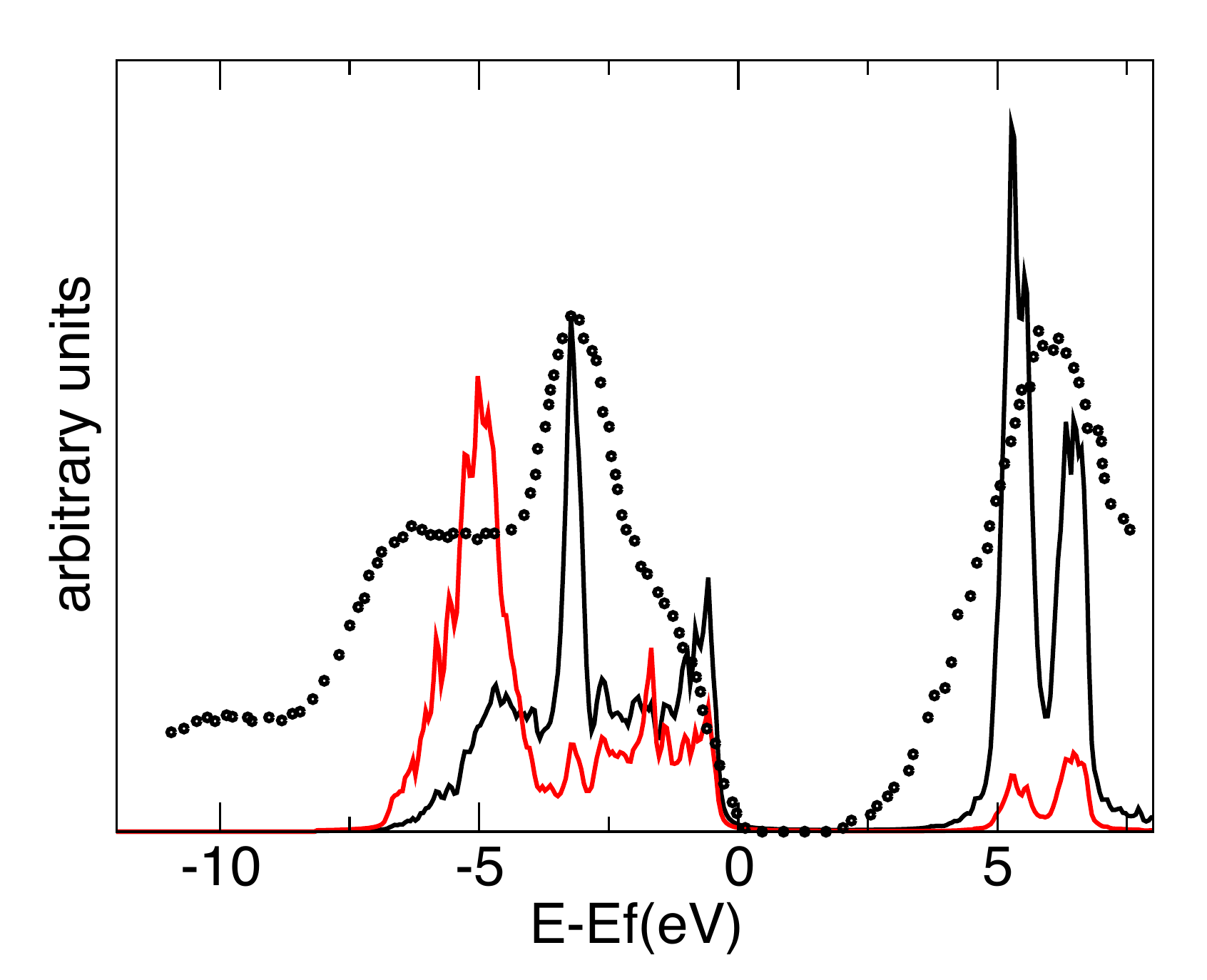}\\
  \caption{(Color in-line) Orbital resolved density of quasi-particle states compared with the experimental XPS and BIS data (circles) of ref. \cite{PhysRevB.44.1530}. Black (red) line is for TM $d$  (Oxygen $p$) orbital contribution. }\label{dosc}
\end{figure}

Figure \ref{dosc} shows  a comparison between the quasiparticle  density of states and  the experimental results of ref. \cite{PhysRevB.44.1530}. We observe that the gap value is well reproduced as well as most of the spectroscopic structures. We do not find evidence of structures below the valence band bottom that are observed in photoemission experiments; this might be due  to the reduced number of excited states that are obtained by the Lanczos procedure. We mention however that the origin of satellites features in MnO has  been somewhat controversial in the literature attributing them either to intrinsic  \cite{PhysRevB.44.1530} or extrinsic effects \cite{Fujimori90}.
A part from the satellite structure our results are comparable with what has been obtained by  Variational Cluster Approximation \cite{Eder} in spite of  a different choice of the cluster , and by a recent DMFT calculation \cite{Dimarco_NiO}. Since these two approaches are either  variationally optimized (VCA) or self-consistent (DMFT), we may identify in our scheme the advantage of giving comparable results by a single shot calculation thanks, we believe, to our cluster choice. Still we are convinced of the importance of variational optimization and our future goal will be to apply it to  our CPT approach.

In conclusion, we have described a method  based on a multi-orbital extension of CPT approach to include on-site interactions in the
description of quasi particle states of real solid systems. The CPT strategy is applied twice, first to identify a  partition of the lattice into non overlapping clusters and secondly to calculate the cluster Green function in terms of two  local ones. This  procedure  has the advantage to replace an unmanageable exact diagonalization by two  separate ones followed by a matrix inversion.
The non-interacting part of the lattice Hamiltonian is described in terms Tight-Binding parameters deduced by a least-square fitting of an ab-initio  single particle band structure, including all the relevant orbitals (no minimal basis set is introduced). To our purposes, since we do not need any real-space  expression of the single particle wavefunctions,  this Tight-Binding parametrization is fully equivalent to a representation in terms of maximally localized Wannier functions. We have applied this method to MnO as a test case and using a single value of Hubbard $U$ we have found a reasonable  agreement with  experimental data and with theoretical results obtained  by different methods. The approach is well suited to treat local correlation in complex materials.

\end{document}